\documentclass[cits]{PoS}

\title{Studying the continuum spectrum of the parsec scale jets by multi-frequency VLBI measurements}

\ShortTitle{Studying the spectra of the parsec scale jets}

\author{\speaker{Tuomas Savolainen}\\
        Max-Planck-Institut f\"ur Radioastronomie, Auf dem H\"ugel 69, 53121
        Bonn, Germany\\
        E-mail: \email{tsavolainen@mpifr-bonn.mpg.de}}

\author{Kaj Wiik, Esko Valtaoja\\
        Tuorla Observatory, Dept. of Physics and Astronomy, University of
        Turku, V\"ais\"al\"antie 20, 21500 Piikki\"o, Finland\\
        E-mail: \email{kaj.wiik@iki.fi}, \email{valtaoja@utu.fi}}

\author{Merja Tornikoski\\
        Mets\"ahovi Radio Observatory, Helsinki Unversity of Technology, 
        Mets\"ahovintie 114, 02540 Kylm\"al\"a, Finland\\
        E-mail: \email{merja.tornikoski@hut.fi}}

      \abstract{Multi-frequency VLBI observations allow studies of the
        continuum spectrum in the different parts of the parsec scale jets of
        AGN, providing information on the physical properties of the plasma
        and the magnetic fields in them. Since VLBI networks cannot be scaled,
        the range of spatial frequencies observed differs significantly
        between the different observing frequencies, which makes it difficult
        to obtain a broadband spectrum of the individual emission features in
        the jet. In this paper we discuss a model-fitting based spectral
        extraction method, which can significantly relieve this problem. The
        method uses {\it a priori} knowledge of the source structure, measured
        at high frequencies, to allow at lower frequencies the derivation of
        the sizes and flux densities of even those emission features that have
        mutual separations significantly less than the Rayleigh limit at the
        given frequency. We have successfully used this method in the analysis
        of 5-86\,GHz VLBA data of 3C\,273. The spectra and sizes of several
        individual jet features were measured, thus allowing derivation of the
        magnetic flux density and the energy density of the relativistic
        electrons in the different parts of the jet. We discuss the results,
        which include e.g. a detection of a strong gradient in the magnetic
        field across the jet of 3C\,273.}

      \FullConference{The 9th European VLBI Network Symposium on The role of
        VLBI in the Golden Age for Radio Astronomy and EVN Users Meeting\\
        September 23-26, 2008\\
        Bologna, Italy}

\begin{document}

\section{Introduction}

Despite the large steps that have been taken in our understanding of the
nature of relativistic outflows in active galactic nuclei during the last
three decades, we still remain ignorant of many of the key issues of the
subject: details of the launching mechanism, matter content and the origin of
the high energy emission are among the hitherto poorly understood aspects of
relativistic jets \cite{mar06}. One of the reasons for slow progress in these
questions is our poor ability to measure some of the key physical parameters
of jets.

There are basically two ways to determine the energy of the relativistic
particles in jets: one is to assume equipartition between the magnetic and
particle energy densities, the other is to measure the size and spectral
turnover of the synchrotron self-absorbed emission features. The latter gives
magnetic flux density and the normalization factor of the electron energy
distribution separately, but it requires a measurement of the 2-D spectrum of
parsec scale jets using multi-frequency VLBI data \cite{mar87}. Such
measurements are possible, yet challenging. In this paper, we describe a new
method for the extraction of the spectral information from a multi-frequency
VLBI data set and apply this method to a set of six-frequency VLBA
observations of the quasar 3C\,273 ($z$=0.158). Throughout the paper, we use a cosmology
with $H_0$ = 71 km s$^{-1}$ Mpc$^{-1}$, $\Omega_M$ = 0.27, and $\Omega
_\Lambda$ = 0.73. This corresponds to a linear scale of 2.7 pc mas$^{-1}$ for
3C\,273.

\section{Model-fitting based spectral extraction method}

\begin{figure}
\centering
\includegraphics[width=0.75\textwidth]{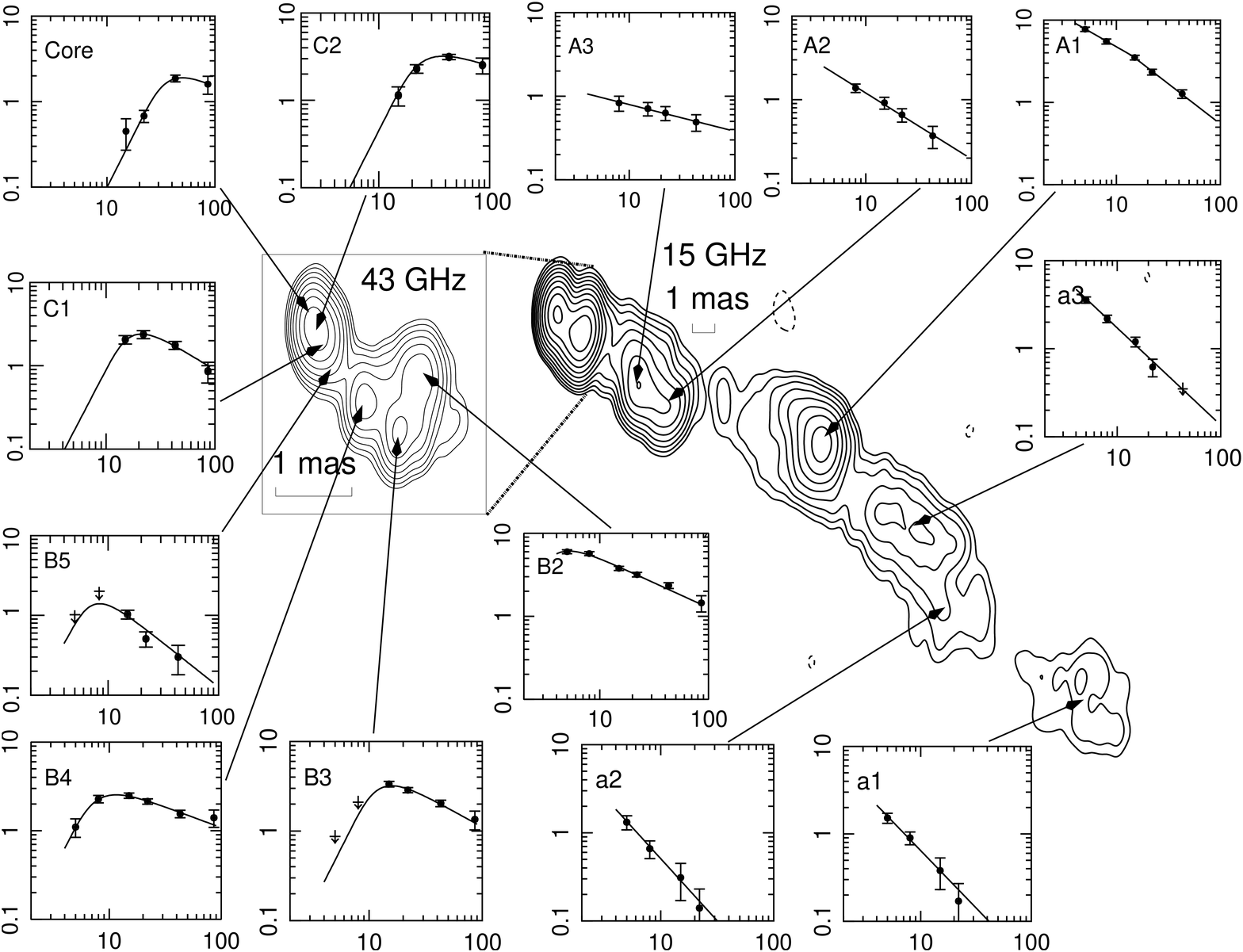}
\caption{Radio spectra of the emission features in the parsec scale jet of
  3C\,273 observed with the VLBA on February 28, 2003. The figure shows also
  Stokes I maps at 15 and 43\,GHz. In the spectral plots, axes are frequency
  in GHz and flux density in Jy. Downward arrows represent upper limits.}
\label{map}
\end{figure}

There are several complications in the analysis of multi-frequency VLBI
observations of extragalactic jets if the aim is to measure a 2-D spectrum
of the jet \cite{lob98}. Since the target sources are typically variable,
measurements at different frequencies need to be
(quasi)simultaneous. Fortunately, this is possible with the VLBA due to its
frequency agility. Also, the flux density calibration should be accurate to
5-10\%. This is typically the case with the VLBA at frequencies $\le 22$\,GHz,
but at higher frequencies special care is required. Registering the images at
different frequencies can be tricky, but there are methods to correct
misalignments between the maps to an acceptable precision \cite{cro08}. The
most severe limiting factor, however, stems from the fact that VLBI networks
are not reconfigurable like e.g. VLA. Therefore, the range of probed spatial
frequencies differs significantly at different observing frequencies. When
comparing VLBI data at two or more observing frequencies, the traditional
solution has been to match the $(u,v)$ coverages either by discarding the data
of the long $(u,v)$ spacings at high frequencies or by tapering the data to a
common resolution. This has the consequence that a significant amount of data
are not used and much of the attainable angular resolution is lost. A broad
frequency coverage, which is needed to reliably measure the synchrotron
self-absorption turnover, exacerbates the problem.

We propose a model-fitting based spectral extraction method to alleviate the
above-described problem of the uneven $(u,v)$ coverage. The idea is to use
{\it a priori} knowledge of the source structure, measured at high
frequencies, to derive, at lower frequencies, the sizes and flux densities of
even those emission features that have mutual separations significantly less
than the beam size at the given frequency. The model of the source structure
at high frequency is taken as a template of the source's brightness
distribution, and the visibility data at lower frequency is fitted with this
template. If the data has high enough signal-to-noise ratio, it is possible to
fit structures that are smaller than the Rayleigh limit \cite{kov05}. This is
equivalent to an out-of-band extrapolation of spatial frequencies, which is
the basis of so-called ``super-resolution'' techniques.

\begin{figure}
\centering
\includegraphics[width=0.6\textwidth]{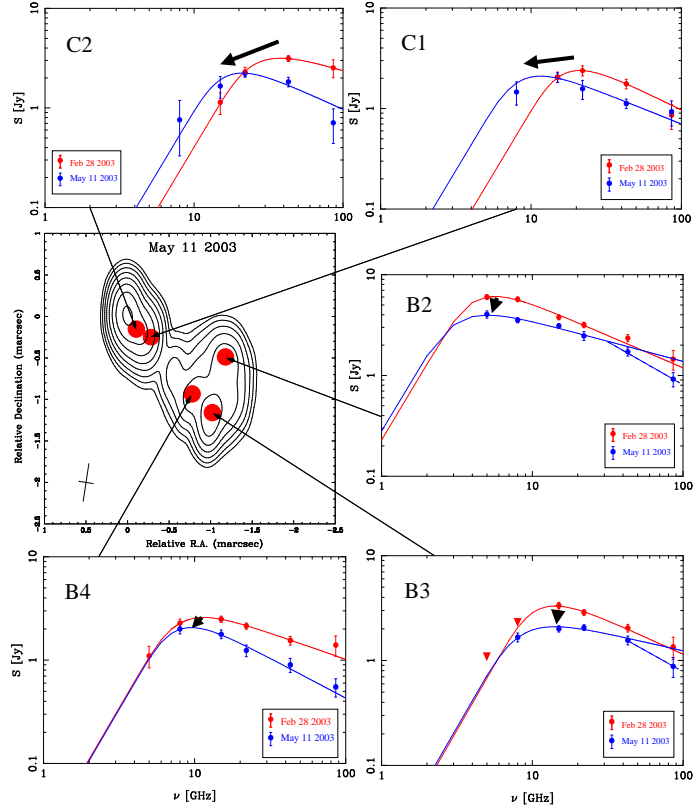}
\caption{Evolution of the spectrum for five emission components in the inner
  part of the jet between February (red) and May (blue) 2003. The component
  positions are marked in the 43\,GHz map from May 2003. Black arrows
  highlight the evolution of the spectral turnover. In the May data,
  components B2 and B3 seem to have a break in their optically thin part of
  the spectrum, which is marked with a dashed line.}
\label{spec_evo}
\end{figure}

In practice, our method can be described as follows (see also \cite{sav08}):
The visibility data at the frequency which gives the best angular resolution
with good SNR, is fitted with a model consisting of a small number of
two-dimensional Gaussian components. This model is our brightness distribution
template, which we use to fit the visibility data at the next lower
frequency. Before fitting, we align the model with the lower frequency data by
assuming that optically thin features in the jet have frequency-independent
positions. This is done in two steps: first-order alignment is done by using
the cross-correlation method of \cite{cro08}, and fine-tuning is done by
moving the whole model around this position in small steps and fitting the
component flux densities at each step while keeping the mutual separations of
the components fixed. The best alignment is found by comparing the post-fit
merit functions. The next step is then to inspect the residual map and add new
components to the model if significant emission, which is not accounted for by
the transfered model, has emerged. Also, all the components that have mutual
separations of less than 1/5 of the beam size are merged into one
component. This is to ensure that we do not extrapolate in spatial frequencies
beyond the $(u,v)$ radius that corresponds to the typical positional accuracy
of the VLBA data. With these modifications to the model, one more round of
fitting is run with component flux densities and sizes as free parameters. The
obtained model is now the brightness distribution template for the next lower
frequency and the above steps are repeated.

Finally, we apply another piece of {\it a priori} knowledge by assuming that
the angular size of the component varies smoothly over the frequencies. The
sizes are inspected as a function of frequency and fitted with either a
constant value or a power-law. Then component sizes in the models are fixed to
the values obtained from these fits and the last round of model-fitting is run
with the component flux densities as the only free parameters. This gives the
final spectra. The main error sources in the final spectra are: 1)
inaccuracies in the alignment, 2) uncertainties due to spatial frequency
extrapolation, and 3) errors in the calibration of the flux density scale. The
effect of (1) can be estimated when the alignment is fine-tuned, (2) has to be
addressed with Monte Carlo simulations, and (3) is best dealt by calibrating
with a single-dish telescope.

\section{Spectrum of the parsec scale jet of 3C\,273}

\begin{figure}
\includegraphics[width=0.49\textwidth]{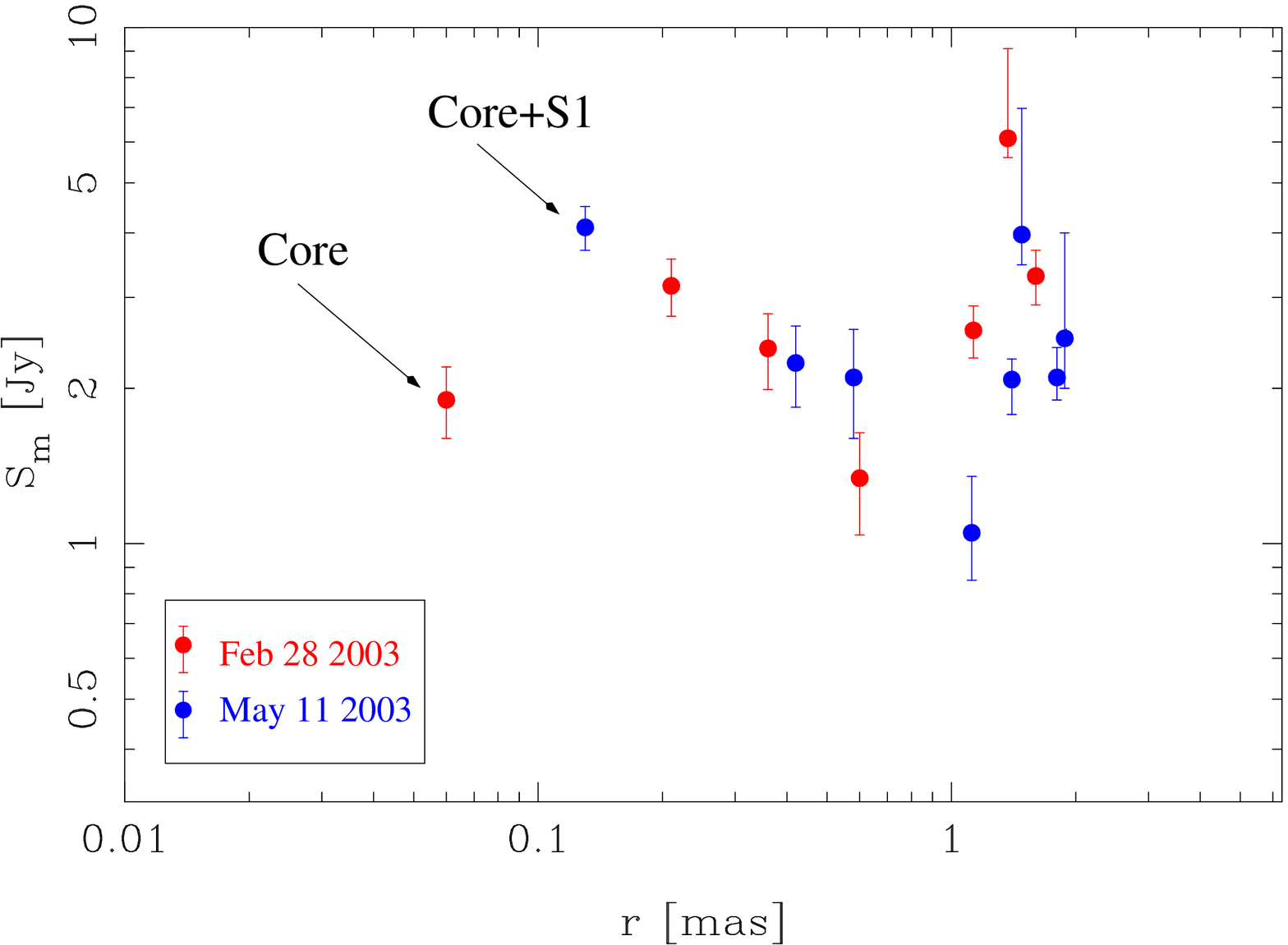}
\includegraphics[width=0.49\textwidth]{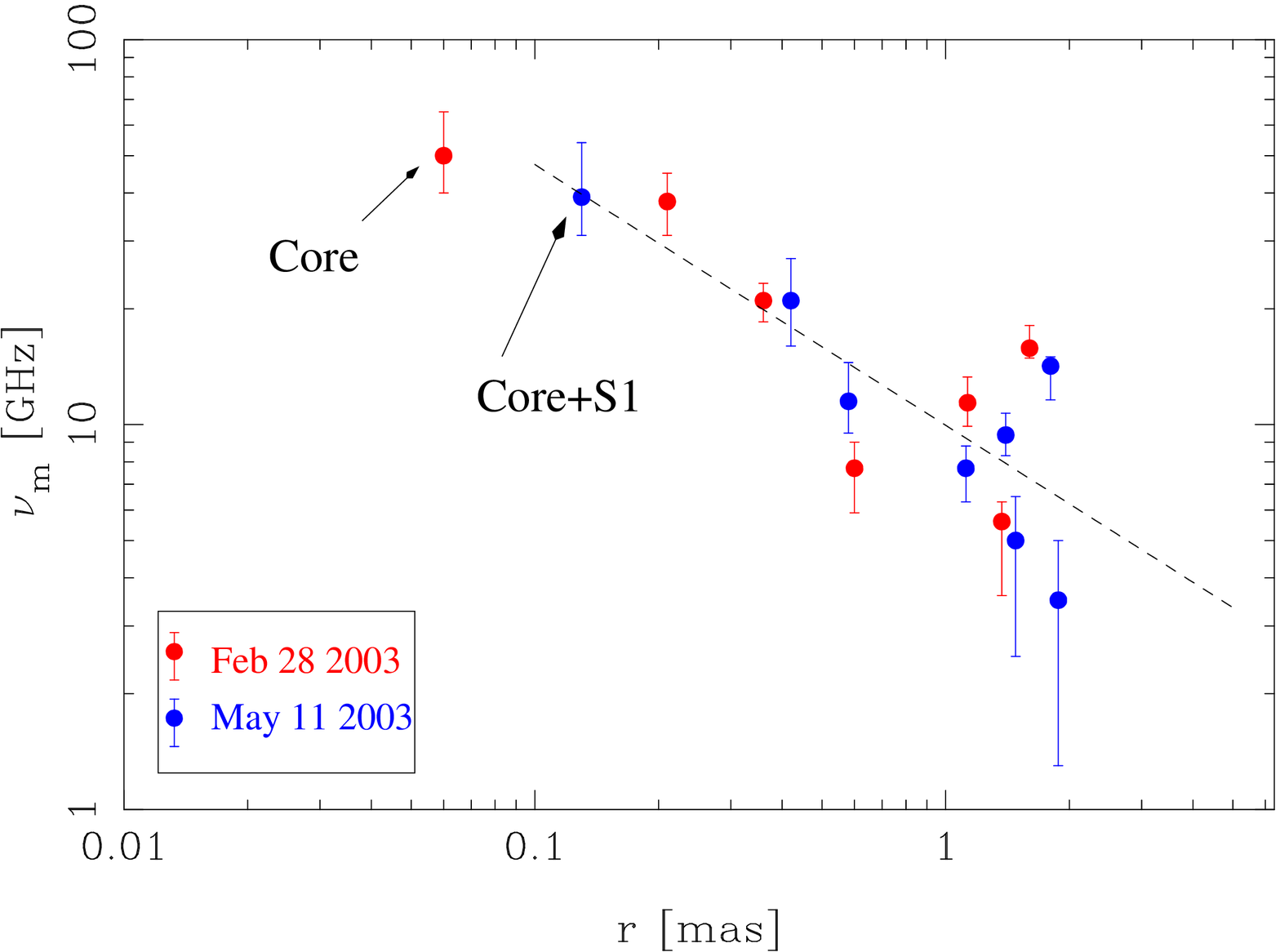}
\caption{Peak flux density (left) and peak frequency (right) of the
  synchrotron spectrum as a function of distance from the core. The dashed
  line corresponds to the power-law function that best fits $\nu_\mathrm{m}$
  of the non-core components. The spectra of the core and the apparently
  stationary component S1 at $r=0.15$\,mas were combined in the May data. See
  \cite{sav08} for a discussion about the choice of the zero point for $r$.}
\label{peak}
\end{figure}

\begin{figure}
\includegraphics[width=0.49\textwidth]{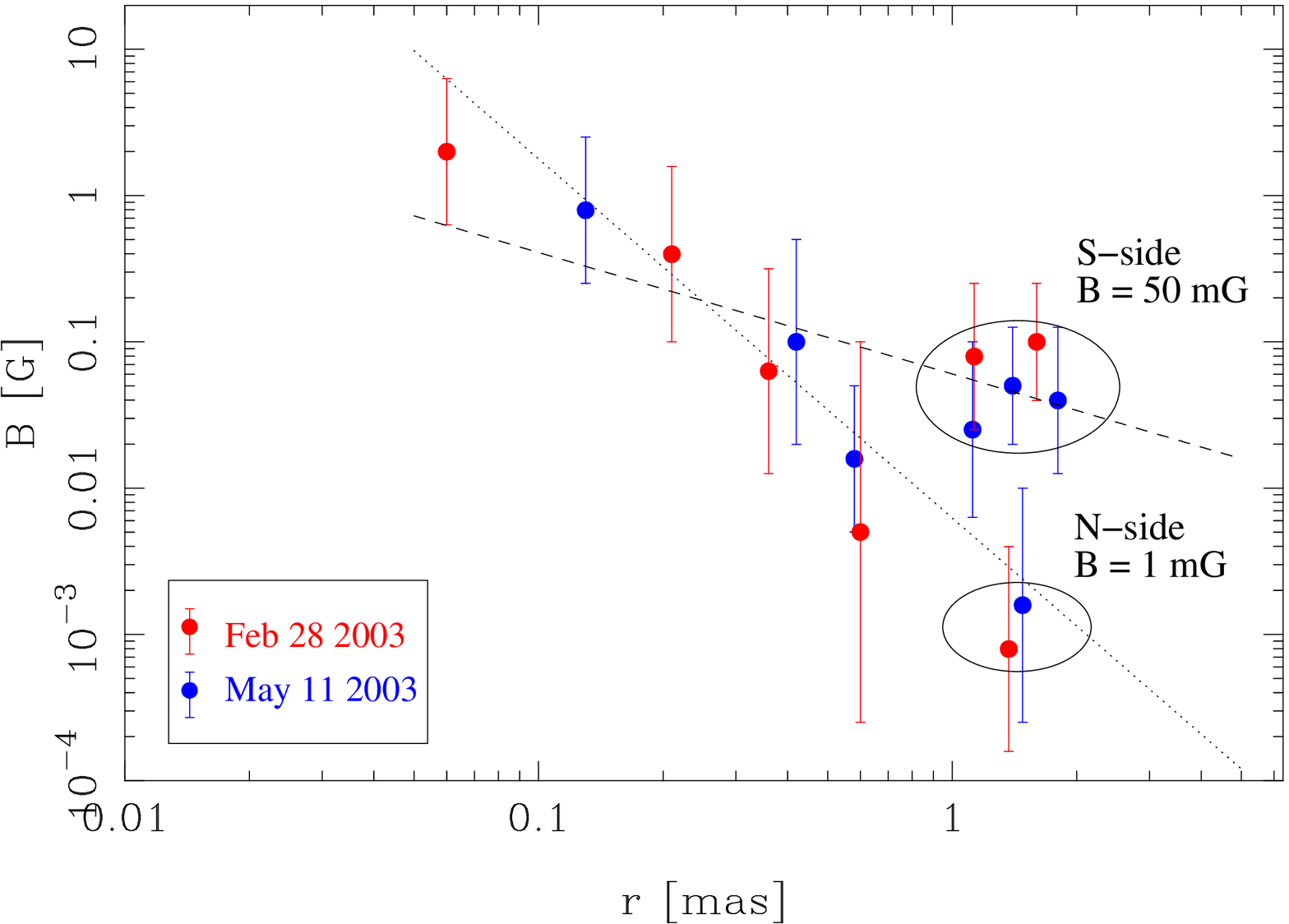}
\includegraphics[width=0.49\textwidth]{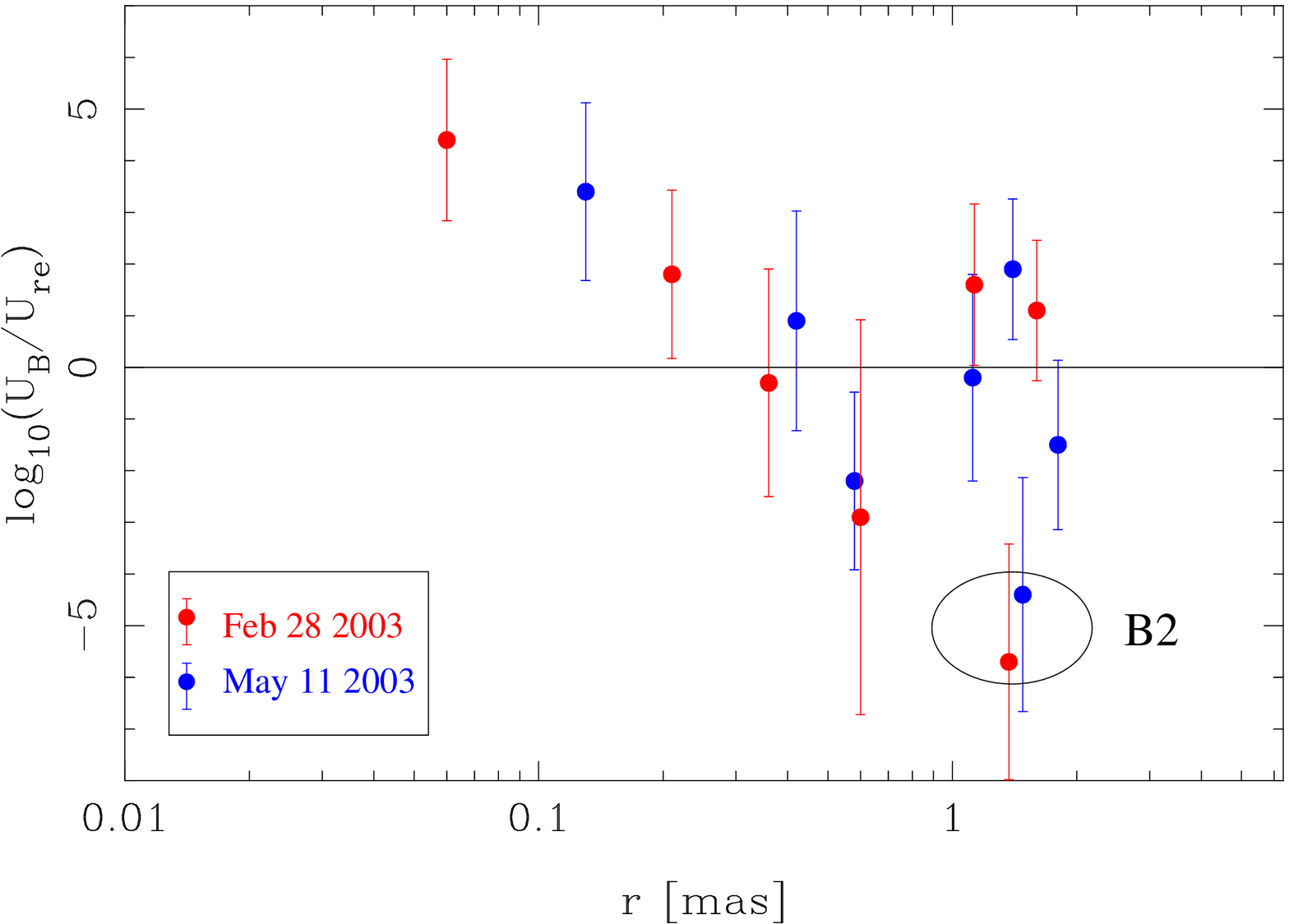}
\caption{Left: Magnetic flux density of the jet components as a function of
  distance from the core. The dashed and dotted lines show examples of
  power-law functions $B\propto r^b$ with $b=-1$ and $b=-2$, respectively. Right: Ratio
  of the magnetic and (radiating) particle energy densities of the components
  as a function of $r$.}
\label{B_N}
\end{figure}

To illustrate the feasibility of the method described in Sect.~2, we present
results from the spectral analysis of the multi-frequency VLBA observations of
3C\,273 that were carried out at five epochs in 2003 \cite{sav06,sav08}. At
every epoch, the source was observed at six frequencies (5.0, 8.4, 15.3, 22.2,
43.2 and 86.2 GHz) with individual scans at different frequencies interleaved
to obtain practically simultaneous multi-frequency data set. An example of the
obtained component spectra are shown in Fig.~\ref{map}. As can be seen in the
figure, the components located within 2\,mas from the core have a spectral
turnover in the observed frequency range.  Therefore, we could fit them with a
function describing self-absorbed synchrotron spectrum from an ensemble of
electrons with a power-law energy distribution in a homogeneous magnetic
field, and derive the turnover frequency, $\nu_\mathrm{m}$, the flux density
at the turnover, $S_\mathrm{m}$, and the optically thin spectral index
$\alpha$. Fig.~\ref{spec_evo} shows a comparison between the spectra observed
at two consecutive epochs for some of the components. It is striking how
consistent the spectra and their evolution are: $\nu_\mathrm{m}$ and
$S_\mathrm{m}$ decrease with time as expected. This gives confidence in the
reliability of the applied spectral extraction method. When more epochs are
analyzed, the spectral evolution can be compared with the predictions of the
shock-in-jet models (see \cite{mar06} and references therein).

The turnover frequencies and flux densities are plotted against the distance
from the core in Fig.~\ref{peak}. The peak frequency decreases with the
distance as $\nu_\mathrm{m} \propto r^{-0.7\pm0.1}$ while the peak flux
density does not seem to have any clear dependence on $r$. This neatly
confirms the compound nature of the flat radio-mm spectrum of 3C\,273.

Since we had measured $S_\mathrm{m}$, $\nu_\mathrm{m}$, and $\alpha$ as well
as the sizes and Doppler factors (see \cite{sav06}) of the components within
2\,mas from the core, we were able to estimate their magnetic flux density,
$B$, and the normalization factor of their electron energy distribution, $N_0$
(without having to assume equipartition) \cite{mar87}. The left panel of
Fig.~\ref{B_N} shows $B$ as a function of $r$. The core has magnetic flux
density of $\sim 1$\,Gauss, which is compatible with the values derived from
infrared and optical variability \cite{cou88}. At $\sim1.5$\,mas from the
core, there is a large discrepancy in $B$ between the southern side
(components B3 and B4; $B\approx50$\,mG) and the northern side of the jet
(component B2; $B\approx1$\,mG). The strong gradient in the magnetic field
density is coincident with the transverse velocity structure found in
\cite{sav06}: the components on the southern side of the jet have higher bulk
velocity than components on the northern side. This may indicate a structured
jet.

Based on the derived values of $B$ and $N_0$, we calculated the ratio between
magnetic and particle energy densities in the components. The right panel of
Fig.~\ref{B_N} shows the results. The uncertainties are large, but some trends
can be seen: the jet starts as magnetically dominated (mm-core) but quickly
moves towards equipartition. Beyond the core, there is one component that is
clearly out of equipartition: B2 is strongly particle dominated. Taking into
account the slow velocity, weak magnetic field and position of component B2,
we possibly see a mixing region where turbulence introduced by velocity shear
converts kinetic and/or magnetic flux into electron energy.

\acknowledgments 

TS is a research fellow of the Alexander von Humboldt Foundation. This work
was also partially supported by the Academy of Finland grant 120516. The VLBA
is a facility of the National Radio Astronomy Observatory, operated by
Associated Universities, Inc., under cooperative agreement with the
U.S. National Science Foundation.

\end{document}